\documentclass[10pt]{article}
\usepackage{graphicx}
\usepackage{amsmath}
\usepackage{amssymb}
\usepackage{caption2}
\begin{document}
\begin{center}
{\large {\bf \sc{  Analysis of the vector hidden-charm-hidden-strange tetraquark  states with implicit P-waves via the QCD sum rules }}} \\[2mm]
Zhi-Gang  Wang \footnote{E-mail: zgwang@aliyun.com.  }     \\
 Department of Physics, North China Electric Power University, Baoding 071003, P. R. China
\end{center}

\begin{abstract}
We take the scalar, pseudoscalar, axialvector, vector and tensor diquarks  as the basic building blocks  to construct  the  four-quark currents with implicit P-waves,  and investigate  the  hidden-charm-hidden-strange tetraquark states with the $J^{PC}=1^{--}$ and $1^{-+}$ via the QCD sum rules in a comprehensive and consistent way,  and revisit the assignments of the  $X/Y$ states, especially the $Y(4500)$, $X(4630)$, $Y(4660)$, $Y(4710)$ and $Y(4790)$, in the   tetraquark  picture.
\end{abstract}



\section{Introduction}
In recent years, a number of  charmonium-like states have been observed, they  cannot be accommodated suitably   in the traditional  quark model. In this work, we will focus on the $Y$ states and tetraquark scenario.
The $Y(4260)$ observed by the BaBar collaboration \cite{BaBar4260-0506}, the $Y(4220)$, $Y(4390)$ and $Y(4320)$ observed  by the BESIII collaboration  \cite{BES-Y4390,BES-Y4220-Y4320}, and the $Y(4360)$, $Y(4660)$ and  $Y(4630)$ observed by the
  Belle collaboration \cite{Belle4660-0707-1,Belle4660-0707-2,Belle4630-0807} are excellent candidates for the vector tetraquark states.

In 2022, the BESIII collaboration observed two resonant structures in the $K^+K^-J/\psi$ invariant mass spectrum, one is the $Y(4230)$ and the other is the $Y(4500)$, which was  observed for the first time with the Breit-Wigner  mass and width $4484.7\pm 13.3\pm 24.1\,\rm{MeV}$ and $111.1\pm 30.1\pm 15.2\, \rm{MeV}$, respectively \cite{BESIII-KK-4500}.

In 2023, the BESIII collaboration observed three enhancements in the  $ D^{*-}D^{*0}\pi^+$ invariant mass spectrum, the first and third resonances are the $Y(4230)$ and $Y(4660)$, respectively, while the second resonance has the Breit-Wigner  mass and width  $4469.1\pm26.2\pm3.6\,\rm{MeV}$ and  $246.3\pm 36.7\pm 9.4\,\rm{MeV}$, respectively, and is roughly compatible with the $Y(4500)$ \cite{X4500-BESIII}.

Also in 2023, the BESIII collaboration observed three resonance structures in the  $ D_{s}^{\ast+}D_{s}^{\ast-}$ invariant mass spectrum, the two significant structures are consistent  with the $\psi(4160)$ and $\psi(4415)$, respectively,  while the third structure is new, and has the Breit-Wigner  mass and width
 $4793.3\pm7.5\,\rm{MeV}$ and $27.1\pm7.0\,\rm{MeV}$, respectively, therefore is named as $Y(4790)$ \cite{Y4790-BESIII}.

Also in 2023,  the BESIII collaboration observed a new resonance, $Y(4710)$, in the $K^+K^-J/\psi$ invariant mass spectrum with a significance over $5\sigma$, the measured  Breit-Wigner  mass and width are  $ 4708_{-15}^{+17}\pm21\, \rm{MeV}$  and $ 126_{-23}^{+27}\pm30\, \rm{MeV}$, respectively \cite{Y4710-BESIII}.

In the picture of tetraquark states and in the theoretical framework of the QCD sum rules, we can construct the color antitriplet-triplet type, sextet-antisextet type,  singlet-singlet type and  octet-octet type  configurations to explore   the properties of the $Y$ states, such as the masses and strong decays \cite{Nielsen-review}. We usually take the diquarks in color antitriplet  $\bar{3}_c$   as the elementary  constituents, as one-gluon exchange leads to  attractive interactions in this channel \cite{One-gluon-1,One-gluon-2}.

The heavy-light diquarks $\varepsilon^{abc}q^{T}_b C\Gamma Q_c$ in color antitriplet  $\bar{3}_c$ have  five  structures, where the $a$, $b$ and $c$ are color indexes, and $C\Gamma=C\gamma_5$, $C$, $C\gamma_\mu \gamma_5$,  $C\gamma_\mu $ and $C\sigma_{\mu\nu}/C\sigma_{\mu\nu}\gamma_5$ for the scalar ($S$), pseudoscalar ($P$), vector ($V$), axialvector ($A$)  and  tensor ($T$) diquarks, respectively.
The $C\gamma_5$ and $C\gamma_\mu$ type diquarks have the spin-parity $J^P=0^+$ and $1^+$, respectively \cite{WZG-diqaurk-cq}, while the $C$ and $C\gamma_\mu\gamma_5$ type diquarks have the spin-parity $J^P=0^-$ and $1^-$, respectively, the implicit
  P-waves   are embodied in the underlined  $\gamma_5$ in the  $C\gamma_5 \underline{\gamma_5} $ and $C\gamma_\mu \underline{\gamma_5} $.
  Under parity transform $\widehat{P}$, the tensor diquarks have the following  properties,
\begin{eqnarray}
\widehat{P}\varepsilon^{abc}q^{T}_b(x)C\sigma_{jk}\gamma_5Q_c(x)\widehat{P}^{-1}&=&
+\varepsilon^{abc}q^{T}_b(\tilde{x})C\sigma_{jk}\gamma_5Q_c(\tilde{x}) \, , \nonumber\\
\widehat{P}\varepsilon^{abc}q^{T}_b(x)C\sigma_{0j}\gamma_5Q_c(x)\widehat{P}^{-1}&=&
-\varepsilon^{abc}q^{T}_b(\tilde{x})C\sigma_{0j}\gamma_5Q_c(\tilde{x}) \, , \nonumber\\
\widehat{P}\varepsilon^{abc}q^{T}_b(x)C\sigma_{0j} Q_c(x)\widehat{P}^{-1}&=&+\varepsilon^{abc}q^{T}_b(\tilde{x})C\sigma_{0j} Q_c(\tilde{x}) \, , \nonumber\\
\widehat{P}\varepsilon^{abc}q^{T}_b(x)C\sigma_{jk} Q_c(x)\widehat{P}^{-1}&=&-\varepsilon^{abc}q^{T}_b(\tilde{x})C\sigma_{jk} Q_c(\tilde{x})\, ,
\end{eqnarray}
where $j$, $k=1$, $2$, $3$, the four vectors $x^\mu=(t,\vec{x})$ and $\tilde{x}^\mu=(t,-\vec{x})$.
 We can see explicitly that the tensor diquarks have both the $J^P=1^+$ and $1^-$ components, and we project out the $J^P=1^+$ and $1^-$ components explicitly by multiplying them with suitable operators when constructing the interpolating currents, and represent  them as $\widetilde{A}$ and $\widetilde{V}$, respectively.

Therefore, we can take the $S$, $P$, $V$, $A$, $\widetilde{A}$ and $\widetilde{V}$ diquarks as  the elementary building blocks  to construct the vector tetraquark states (or four-quark currents) with  an implicit P-wave, or introduce an explicit P-wave between the S-wave diquark pairs to construct the vector tetraquark states (or four-quark currents).

In Refs.\cite{Vector-Tetra-WZG-P-wave-1,Vector-Tetra-WZG-P-wave}, we introduce a relative P-wave between the diquark pair explicitly to construct the local four-quark currents,  and explore  the  vector tetraquark states in the framework of  the QCD sum rules systematically, and obtain the lowest vector tetraquark masses up to now.  The  predictions support  assigning the
 $Y(4220/4260)$,  $Y(4320/4360)$, $Y(4390)$ and $Z(4250)$ as  the lowest vector tetraquark   states with a relative P-wave between the diquark pair.

While in the  type-II diquark model \cite{Maiani-II-type},  L. Maiani et al assign  the $Y(4008)$, $Y(4260)$, $Y(4290/4220)$  and $Y(4630)$ as   the four ground  states with  $L=1$ based on the effective  Hamiltonian with the spin-spin and spin-orbit  interactions but neglecting the $q-\bar{q}$ spin-spin interactions. In Ref.\cite{Ali-Maiani-Y}, A. Ali et al incorporate the dominant spin-spin, spin-orbit and tensor interactions, and observe that the preferred  assignments of the ground states with  $L=1$ are the $Y(4220)$, $Y(4330)$, $Y(4390)$, $Y(4660)$, however, the mass-splitting effects among the multiplet are too large.

In  Ref.\cite{WZG-tetra-vector-NPB},  we take the $S$, $P$, $V$, $A$, $\widetilde{A}$ and $\widetilde{V}$    diquarks  as the elementary building blocks  to construct  four-quark  currents with an implicit P-wave,  and explore   the  hidden-charm tetraquark states with the $J^{PC}=1^{--}$ and $1^{-+}$ in the framework of the QCD sum rules comprehensively,  and revisit the assignments of the  $Y$ states in the  picture  of tetraquark  states.

In fact, all the predictions depend heavily on the input parameters, for example, the same interpolating currents result in quite different assignments of the $Y(4660)$ \cite{Nielsen-4260-4660,ChenZhu,WangY4360Y4660-1803}, let alone different interpolating currents \cite{WZG-tetra-vector-NPB,Nielsen-4260-4660,ChenZhu,WangY4360Y4660-1803,Wang-tetra-formula,Wang-CTP-4660,
Nielsen-4660-mole,Azizi-4660,WangEPJC-1601,
ZhangHuang-PRD}, see Table \ref{Y4660-Ref}.
A comprehensive analysis with the same input parameters and same (or uniform) treatments is necessary, in this direction, beyond the achievements acquired in Refs.\cite{Vector-Tetra-WZG-P-wave,WZG-tetra-vector-NPB}, we also have explored  the hidden-charm tetraquark states with the $J^{PC}=0^{++}$,  $0^{-+}$, $0^{--}$,  $1^{+-}$, $2^{++}$ \cite{WZG-HC-spectrum-PRD,WZG-tetra-psedo-NPB}, hidden-bottom tetraquark states with the $J^{PC}=0^{++}$, $1^{+-}$, $2^{++}$ \cite{WZG-HB-spectrum-EPJC}, hidden-charm molecular states with the $J^{PC}=0^{++}$, $1^{+-}$, $2^{++}$ \cite{WZG-mole-IJMPA}, doubly-charmed tetraquark (molecular) states with the $J^{P}=0^{+}$, $1^{+}$, $2^{+}$ \cite{WZG-tetra-cc-EPJC} (\cite{WZG-XQ-mole-EPJA}).  In our previous works \cite{Vector-Tetra-WZG-P-wave-1,Vector-Tetra-WZG-P-wave,WZG-tetra-vector-NPB,WZG-HC-spectrum-PRD,WZG-tetra-psedo-NPB,
WZG-HB-spectrum-EPJC,WZG-mole-IJMPA,WZG-tetra-cc-EPJC,WZG-XQ-mole-EPJA}, we resort to the energy scale formula  or modified energy scale formula  to choose the best energy scales of the QCD spectral densities to enhance  the pole contributions and improve the convergent behaviors of the operator product expansion \cite{Wang-tetra-formula}, it is the unique feature of our works.

\begin{table}
\begin{center}
\begin{tabular}{|c|c|c|c|c|c|c|c|}\hline\hline
           &Structures                                                      &OPE\,(No)    &mass(GeV)      &References   \\ \hline

$Y(4660)$  &$\psi^\prime f_0(980)$                                          &$10$         &$4.71$         &\cite{Wang-CTP-4660}  \\
$Y(4660)$  &$\psi^\prime f_0(980)$                                          &$6$          &$4.67$         &\cite{Nielsen-4660-mole}  \\ \hline

$Y(4660)$  &$[sc]_S[\overline{sc}]_V+[sc]_V[\overline{sc}]_S$               &$8\,(7)$     &$4.65$         &\cite{Nielsen-4260-4660}  \\
$Y(4660)$  &$[sc]_S[\overline{sc}]_V+[sc]_V[\overline{sc}]_S$               &$10$         &$4.68$         &\cite{Azizi-4660}  \\  \hline

$Y(4660)$  &$[qc]_S[\overline{qc}]_V+[qc]_V[\overline{qc}]_S$               &$8\,(7)$     &$4.64$         &\cite{ChenZhu}  \\
$Y(4360)$  &$[qc]_S[\overline{qc}]_V+[qc]_V[\overline{qc}]_S$              &$10$         &$4.34$         &\cite{WangY4360Y4660-1803}  \\ \hline

$Y(4660)$  &$[sc]_P[\overline{sc}]_A-[sc]_A[\overline{sc}]_P$               &$10$         &$4.70$         &\cite{Wang-tetra-formula}  \\
$Y(4660)$  &$[sc]_P[\overline{sc}]_A-[sc]_A[\overline{sc}]_P$               &$10$         &$4.66$         &\cite{WangY4360Y4660-1803}  \\ \hline

$Y(4660)$  &$[qc]_P[\overline{qc}]_A-[qc]_A[\overline{qc}]_P$               &$10$         &$4.66$         &\cite{WZG-tetra-vector-NPB}   \\

$Y(4660)$  &$[qc]_P[\overline{qc}]_A-[qc]_A[\overline{qc}]_P$               &$10$         &$4.66$         &\cite{Wang-tetra-formula}   \\

$Y(4660)$  &$[qc]_P[\overline{qc}]_A-[qc]_A[\overline{qc}]_P$               &$10$         &$4.59$         &\cite{WangY4360Y4660-1803}  \\  \hline

$Y(4660)$  &$[qc]_A[\overline{qc}]_A$                                       &$10$         &$4.66$         &\cite{WangEPJC-1601}  \\ \hline

$Y(4660)$  &$[qc]_A[\overline{qc}]_A$                                       &$10$         &$4.69$         &\cite{WZG-tetra-vector-NPB}  \\ \hline

$Y(4660)$  &$[sc]_S[\overline{sc}]_S$                                       &$6$          &$4.69$         &\cite{ZhangHuang-PRD}  \\ \hline

$Y(4660)$   &$[sc]_{\tilde{V}}[\overline{sc}]_{A}-[sc]_{A}[\overline{sc}]_{\tilde{V}}$  &$8\,(7)$     &$4.64$         &\cite{ChenZhu}  \\ \hline

$Y(4660)$  &$[sc]_{\tilde{A}}[\overline{sc}]_{V}+[sc]_{V}[\overline{sc}]_{\tilde{A}}$  &$10$  &$4.65$ &This work \\

$Y(4660)$  &$[sc]_{S}[\overline{sc}]_{\tilde{V}}-[sc]_{\tilde{V}}[\overline{sc}]_{S}$  &$10$  &$4.68$ &This work  \\ \hline \hline
\end{tabular}
\end{center}
\caption{ The  masses from the QCD sum rules with different quark structures, where the OPE denotes  truncations of  the operator product expansion up to the vacuum condensates of dimension $n$, the No denotes the vacuum condensates of dimension $n^\prime$ are not included.   } \label{Y4660-Ref}
\end{table}

 In this work, we  take the $S$, $P$, $V$, $A$, $\widetilde{A}$ and $\widetilde{V}$ diquarks as  the elementary  building blocks  to construct the local four-quark currents with an implicit P-wave to realize the negative parity, and   investigate    the  hidden-charm-hidden-strange tetraquark states with the $J^{PC}=1^{--}$ and $1^{-+}$ in a comprehensive and consistent way. We  take account of the light-flavor $SU(3)$-breaking effects and try to make possible assignments of the $X/Y$ states combined with Ref.\cite{WZG-tetra-vector-NPB} in a consistent way.

The article is arranged as follows:  we obtain the QCD sum rules for the vector hidden-charm-hidden-strange  tetraquark states in section 2; in section 3, we   present the numerical results and discussions; section 4 is reserved for our conclusion.

\section{QCD sum rules for  the  vector hidden-charm-hidden-strange  tetraquark states}
We  write down  the two-point correlation functions  $\Pi_{\mu\nu}(p)$ and $\Pi_{\mu\nu\alpha\beta}(p)$ firstly,
\begin{eqnarray}\label{CF-Pi}
\Pi_{\mu\nu}(p)&=&i\int d^4x e^{ip \cdot x} \langle0|T\Big\{J_\mu(x)J_{\nu}^{\dagger}(0)\Big\}|0\rangle \, ,\nonumber\\
\Pi_{\mu\nu\alpha\beta}(p)&=&i\int d^4x e^{ip \cdot x} \langle0|T\Big\{J_{\mu\nu}(x)J_{\alpha\beta}^{\dagger}(0)\Big\}|0\rangle \, ,
\end{eqnarray}
where the currents
\begin{eqnarray}
J_\mu(x)&=&J^{PA}_{-,\mu}(x)\, ,\,\, J^{PA}_{+,\mu}(x)\, , \,\,J^{SV}_{-,\mu}(x)\, , \,\, J^{SV}_{+,\mu}(x)\, ,\,\,
J_{-,\mu}^{\widetilde{V}A}(x)\, ,\,\, J_{+,\mu}^{\widetilde{V}A}(x)\, , \,\, J_{-,\mu}^{\widetilde{A}V}(x)\, , \,\, J_{+,\mu}^{\widetilde{A}V}(x)\, , \nonumber\\
J_{\mu\nu}(x)&=&J^{S\widetilde{V}}_{-,\mu\nu}(x)\, , \,\, J^{S\widetilde{V}}_{+,\mu\nu}(x)\, , \,\, J^{P\widetilde{A}}_{-,\mu\nu}(x)\, , \,\,
J^{P\widetilde{A}}_{+,\mu\nu}(x)\, , \, \, J^{AA}_{-,\mu\nu}(x)\, ,
\end{eqnarray}
\begin{eqnarray}\label{PA-SV}
J^{PA}_{-,\mu}(x)&=&\frac{\varepsilon^{ijk}\varepsilon^{imn}}{\sqrt{2}}\Big[s^{T}_j(x)Cc_k(x) \bar{s}_m(x)\gamma_\mu C \bar{c}^{T}_n(x)-s^{T}_j(x)C\gamma_\mu c_k(x)\bar{s}_m(x)C \bar{c}^{T}_n(x) \Big] \, ,\nonumber\\
J^{PA}_{+,\mu}(x)&=&\frac{\varepsilon^{ijk}\varepsilon^{imn}}{\sqrt{2}}\Big[s^{T}_j(x)Cc_k(x) \bar{s}_m(x)\gamma_\mu C \bar{c}^{T}_n(x)+s^{T}_j(x)C\gamma_\mu c_k(x)\bar{s}_m(x)C \bar{c}^{T}_n(x) \Big] \, ,\nonumber\\
J^{SV}_{-,\mu}(x)&=&\frac{\varepsilon^{ijk}\varepsilon^{imn}}{\sqrt{2}}\Big[s^{T}_j(x)C\gamma_5c_k(x) \bar{s}_m(x)\gamma_5\gamma_\mu C \bar{c}^{T}_n(x)+s^{T}_j(x)C\gamma_\mu\gamma_5 c_k(x)\bar{s}_m(x)\gamma_5C \bar{c}^{T}_n(x) \Big] \, ,\nonumber\\
J^{SV}_{+,\mu}(x)&=&\frac{\varepsilon^{ijk}\varepsilon^{imn}}{\sqrt{2}}\Big[s^{T}_j(x)C\gamma_5c_k(x) \bar{s}_m(x)\gamma_5\gamma_\mu C \bar{c}^{T}_n(x)-s^{T}_j(x)C\gamma_\mu\gamma_5 c_k(x)\bar{s}_m(x)\gamma_5C \bar{c}^{T}_n(x) \Big] \, ,\nonumber\\
\end{eqnarray}

\begin{eqnarray}\label{VA-AV}
J_{-,\mu}^{\widetilde{V}A}(x)&=&\frac{\varepsilon^{ijk}\varepsilon^{imn}}{\sqrt{2}}\Big[s^{T}_j(x)C\sigma_{\mu\nu} c_k(x)\bar{s}_m(x)\gamma^\nu C \bar{c}^{T}_n(x)-s^{T}_j(x)C\gamma^\nu c_k(x)\bar{s}_m(x)\sigma_{\mu\nu} C \bar{c}^{T}_n(x) \Big] \, , \nonumber\\
J_{+,\mu}^{\widetilde{V}A}(x)&=&\frac{\varepsilon^{ijk}\varepsilon^{imn}}{\sqrt{2}}\Big[s^{T}_j(x)C\sigma_{\mu\nu} c_k(x)\bar{s}_m(x)\gamma^\nu C \bar{c}^{T}_n(x)+s^{T}_j(x)C\gamma^\nu c_k(x)\bar{s}_m(x)\sigma_{\mu\nu} C \bar{c}^{T}_n(x) \Big] \, , \nonumber\\
J_{-,\mu}^{\widetilde{A}V}(x)&=&\frac{\varepsilon^{ijk}\varepsilon^{imn}}{\sqrt{2}}\Big[s^{T}_j(x)
C\sigma_{\mu\nu}\gamma_5 c_k(x)\bar{s}_m(x)\gamma_5\gamma^\nu C \bar{c}^{T}_n(x)+s^{T}_j(x)C\gamma^\nu\gamma_5 c_k(x)\bar{s}_m(x)\gamma_5\sigma_{\mu\nu} C \bar{c}^{T}_n(x) \Big] \, , \nonumber\\
J_{+,\mu}^{\widetilde{A}V}(x)&=&\frac{\varepsilon^{ijk}\varepsilon^{imn}}{\sqrt{2}}\Big[s^{T}_j(x)
C\sigma_{\mu\nu}\gamma_5 c_k(x)\bar{s}_m(x)\gamma_5\gamma^\nu C \bar{c}^{T}_n(x)-s^{T}_j(x)C\gamma^\nu\gamma_5 c_k(x)\bar{s}_m(x)\gamma_5\sigma_{\mu\nu} C \bar{c}^{T}_n(x) \Big] \, , \nonumber\\
\end{eqnarray}

\begin{eqnarray}\label{SV-PA-T}
J^{S\widetilde{V}}_{-,\mu\nu}(x)&=&\frac{\varepsilon^{ijk}\varepsilon^{imn}}{\sqrt{2}}\Big[s^{T}_j(x)C\gamma_5 c_k(x)  \bar{s}_m(x)\sigma_{\mu\nu} C \bar{c}^{T}_n(x)- s^{T}_j(x)C\sigma_{\mu\nu} c_k(x)  \bar{s}_m(x)\gamma_5 C \bar{c}^{T}_n(x) \Big] \, , \nonumber\\
J^{S\widetilde{V}}_{+,\mu\nu}(x)&=&\frac{\varepsilon^{ijk}\varepsilon^{imn}}{\sqrt{2}}\Big[s^{T}_j(x)C\gamma_5 c_k(x)  \bar{s}_m(x)\sigma_{\mu\nu} C \bar{c}^{T}_n(x)+ s^{T}_j(x)C\sigma_{\mu\nu} c_k(x)  \bar{s}_m(x)\gamma_5 C \bar{c}^{T}_n(x) \Big] \, , \nonumber\\
J^{P\widetilde{A}}_{-,\mu\nu}(x)&=&\frac{\varepsilon^{ijk}\varepsilon^{imn}}{\sqrt{2}}\Big[s^{T}_j(x)C c_k(x)  \bar{s}_m(x)\gamma_5\sigma_{\mu\nu} C \bar{c}^{T}_n(x)- s^{T}_j(x)C\sigma_{\mu\nu}\gamma_5 c_k(x)  \bar{s}_m(x)  C \bar{c}^{T}_n(x) \Big] \, , \nonumber\\
J^{P\widetilde{A}}_{+,\mu\nu}(x)&=&\frac{\varepsilon^{ijk}\varepsilon^{imn}}{\sqrt{2}}\Big[s^{T}_j(x)C c_k(x)  \bar{s}_m(x)\gamma_5\sigma_{\mu\nu} C \bar{c}^{T}_n(x)+ s^{T}_j(x)C\sigma_{\mu\nu}\gamma_5 c_k(x)  \bar{s}_m(x)  C \bar{c}^{T}_n(x) \Big] \, , \nonumber\\
\end{eqnarray}

\begin{eqnarray}\label{AA-J-2}
J^{AA}_{-,\mu\nu}(x)&=&\frac{\varepsilon^{ijk}\varepsilon^{imn}}{\sqrt{2}}\Big[s^{T}_j(x) C\gamma_\mu c_k(x) \bar{s}_m(x) \gamma_\nu C \bar{c}^{T}_n(x)  -s^{T}_j(x) C\gamma_\nu c_k(x) \bar{s}_m(x) \gamma_\mu C \bar{c}^{T}_n(x) \Big] \, ,  \nonumber\\
\end{eqnarray}
 the $i$, $j$, $k$, $m$, $n$ are  color indexes,   the $C$ is the charge conjugation matrix, the subscripts $\pm$ represent  the positive  and negative charge conjugation, respectively, the superscripts $P$, $S$, $A$($\widetilde{A}$) and $V$($\widetilde{V}$) represent  the pseudoscalar, scalar, axialvector and vector diquarks or  antidiquarks, respectively.

 The currents in Eq.\eqref{PA-SV} have one Lorentz index and are made of the $S$, $P$, $A$ and $V$ diquarks. The currents  in Eq.\eqref{VA-AV} also have one Lorentz index but are made of the $A$, $V$, $\widetilde{A}$ and $\widetilde{V}$ diquarks, thus involve the tensor diquarks.
 The currents in Eq.\eqref{SV-PA-T} have two Lorentz indexes which are antisymmetric, and are made of the $S$, $P$, $\widetilde{A}$ and $\widetilde{V}$ diquarks, thus also involve the tensor diquarks. The current in Eq.\eqref{AA-J-2} has two Lorentz indexes which are also antisymmetric but are made of the $A$ diquarks not the $T$ diquarks.

  Under parity transform $\widehat{P}$, the currents $J_\mu(x)$ and $J_{\mu\nu}(x)$ have the  properties,
\begin{eqnarray}\label{J-parity}
\widehat{P} J_\mu(x)\widehat{P}^{-1}&=&+J^\mu(\tilde{x}) \, , \nonumber\\
\widehat{P} \tilde{J}_{\mu\nu}(x)\widehat{P}^{-1}&=&-\tilde{J}^{\mu\nu}(\tilde{x}) \, , \nonumber\\
\widehat{P} J^{AA}_{-,\mu\nu}(x)\widehat{P}^{-1}&=&+J_{AA}^{-,\mu\nu}(\tilde{x}) \, ,
\end{eqnarray}
where $\tilde{J}_{\mu\nu}(x)=J^{S\widetilde{V}}_{-,\mu\nu}(x)$,
$J^{S\widetilde{V}}_{+,\mu\nu}(x)$,
$J^{P\widetilde{A}}_{-,\mu\nu}(x)$,
$J^{P\widetilde{A}}_{+,\mu\nu}(x)$, the coordinates $x^\mu=(t,\vec{x})$ and $\tilde{x}^\mu=(t,-\vec{x})$.
 We rewrite Eq.\eqref{J-parity} more explicitly to make sense of the parity eigenstates,
\begin{eqnarray}
\widehat{P} J_i(x)\widehat{P}^{-1}&=&-J_i(\tilde{x}) \, , \nonumber\\
\widehat{P} \tilde{J}_{ij}(x)\widehat{P}^{-1}&=&-\tilde{J}_{ij}(\tilde{x}) \, , \nonumber\\
\widehat{P} J^{AA}_{-,0i}(x)\widehat{P}^{-1}&=&-J_{AA,0i}^{-}(\tilde{x}) \, ,
\end{eqnarray}
\begin{eqnarray}
\widehat{P} J_0(x)\widehat{P}^{-1}&=&+J_0(\tilde{x}) \, , \nonumber\\
\widehat{P} \tilde{J}_{0i}(x)\widehat{P}^{-1}&=&+\tilde{J}_{0i}(\tilde{x}) \, , \nonumber\\
\widehat{P} J^{AA}_{-,ij}(x)\widehat{P}^{-1}&=&+J_{AA,ij}^{-}(\tilde{x}) \, ,
\end{eqnarray}
where $i$, $j=1$, $2$, $3$. The currents $J_\mu(x)$ and $J_{\mu\nu}(x)$ have both  negative and positive-parity components, which couple potentially to the hidden-charm-hidden-strange tetraquark states with the  negative and positive-parity, respectively, we distinguish their contributions explicitly  by choosing the pertinent tensor structures.

Under charge-conjugation transform $\widehat{C}$, the currents  $J_\mu(x)$ and $J_{\mu\nu}(x)$ have the properties,
\begin{eqnarray}
\widehat{C}J_{\pm,\mu}(x)\widehat{C}^{-1}&=&\pm J_{\pm,\mu}(x)  \, , \nonumber\\
\widehat{C}J_{\pm,\mu\nu}(x)\widehat{C}^{-1}&=&\pm J_{\pm,\mu\nu}(x)  \, ,
\end{eqnarray}
 the interpolating currents have definite charge conjugation, and couple potentially to the tetraquark states with definite charge conjugation.

The current $J^{PA}_{-,\mu}(x)$ has been studied in Refs.\cite{WangY4360Y4660-1803,Wang-tetra-formula},  while  the currents $J^{SV}_{-,\mu}(x)$ and $J^{PA}_{+,\mu}(x)$ have  been studied in Ref.\cite{WangY4360Y4660-1803} and Ref.\cite{Wang-tetra-formula}, respectively. In this work, we update the old calculations \cite{WangY4360Y4660-1803,Wang-tetra-formula}, and extend our previous work on the mass spectrum of the hidden-charm tetraquark states with the $J^{PC}=1^{-\pm}$ \cite{WZG-tetra-vector-NPB} to the mass spectrum of the hidden-charm-hidden-strange tetraquark states with the $J^{PC}=1^{-\pm}$  by taking account of the light flavor $SU(3)$ breaking effects,  and perform a comprehensive and consistent  analysis.

At the hadron side, we  insert  a complete set of  intermediate hadronic states with
the same quantum numbers as the currents  $J_\mu(x)$ and $J_{\mu\nu}(x)$ into the
correlation functions  $\Pi_{\mu\nu}(p)$ and $\Pi_{\mu\nu\alpha\beta}(p)$   to reach  the hadronic representation
\cite{SVZ79-1,SVZ79-2,Reinders85}, and isolate the lowest  hidden-charm-hidden-strange tetraquark contributions,
\begin{eqnarray}
\Pi_{\mu\nu}(p)&=&\frac{\lambda_{Y_{-}}^2}{M_{Y_{-}}^2-p^2}\left( -g_{\mu\nu}+\frac{p_{\mu}p_{\nu}}{p^2}\right) +\cdots \nonumber\\
&=&\Pi_{-}(p^2)\left( -g_{\mu\nu}+\frac{p_{\mu}p_{\nu}}{p^2}\right)+\cdots \, ,\nonumber\\
\widetilde{\Pi}_{\mu\nu\alpha\beta}(p)&=&\frac{\lambda_{Y_{-}}^2}{M_{Y_{-}}^2\left(M_{Y_{-}}^2-p^2\right)}\left(p^2g_{\mu\alpha}g_{\nu\beta} -p^2g_{\mu\beta}g_{\nu\alpha} -g_{\mu\alpha}p_{\nu}p_{\beta}-g_{\nu\beta}p_{\mu}p_{\alpha}+g_{\mu\beta}p_{\nu}p_{\alpha}+g_{\nu\alpha}p_{\mu}p_{\beta}\right) \nonumber\\
&&+\frac{\lambda_{ Z_{+}}^2}{M_{Z_{+}}^2\left(M_{Z_{+}}^2-p^2\right)}\left( -g_{\mu\alpha}p_{\nu}p_{\beta}-g_{\nu\beta}p_{\mu}p_{\alpha}+g_{\mu\beta}p_{\nu}p_{\alpha}+g_{\nu\alpha}p_{\mu}p_{\beta}\right) +\cdots  \nonumber\\
&=&\widetilde{\Pi}_{-}(p^2)\left(p^2g_{\mu\alpha}g_{\nu\beta} -p^2g_{\mu\beta}g_{\nu\alpha} -g_{\mu\alpha}p_{\nu}p_{\beta}-g_{\nu\beta}p_{\mu}p_{\alpha}+g_{\mu\beta}p_{\nu}p_{\alpha}+g_{\nu\alpha}p_{\mu}p_{\beta}\right) \nonumber\\
&&+\widetilde{\Pi}_{+}(p^2)\left( -g_{\mu\alpha}p_{\nu}p_{\beta}-g_{\nu\beta}p_{\mu}p_{\alpha}+g_{\mu\beta}p_{\nu}p_{\alpha}+g_{\nu\alpha}p_{\mu}p_{\beta}\right) \, ,\nonumber\\
\Pi^{AA}_{\mu\nu\alpha\beta}(p)&=&\frac{\lambda_{ Z_{+}}^2}{M_{Z_{+}}^2\left(M_{Z_{+}}^2-p^2\right)}\left(p^2g_{\mu\alpha}g_{\nu\beta} -p^2g_{\mu\beta}g_{\nu\alpha} -g_{\mu\alpha}p_{\nu}p_{\beta}-g_{\nu\beta}p_{\mu}p_{\alpha}+g_{\mu\beta}p_{\nu}p_{\alpha}+g_{\nu\alpha}p_{\mu}p_{\beta}\right) \nonumber\\
&&+\frac{\lambda_{Y_{ -}}^2}{M_{Y_{-}}^2\left(M_{Y_{-}}^2-p^2\right)}\left( -g_{\mu\alpha}p_{\nu}p_{\beta}-g_{\nu\beta}p_{\mu}p_{\alpha}+g_{\mu\beta}p_{\nu}p_{\alpha}+g_{\nu\alpha}p_{\mu}p_{\beta}\right) +\cdots  \nonumber\\
&=&\widetilde{\Pi}_{+}(p^2)\left(p^2g_{\mu\alpha}g_{\nu\beta} -p^2g_{\mu\beta}g_{\nu\alpha} -g_{\mu\alpha}p_{\nu}p_{\beta}-g_{\nu\beta}p_{\mu}p_{\alpha}+g_{\mu\beta}p_{\nu}p_{\alpha}+g_{\nu\alpha}p_{\mu}p_{\beta}\right) \nonumber\\
&&+\widetilde{\Pi}_{-}(p^2)\left( -g_{\mu\alpha}p_{\nu}p_{\beta}-g_{\nu\beta}p_{\mu}p_{\alpha}+g_{\mu\beta}p_{\nu}p_{\alpha}+g_{\nu\alpha}p_{\mu}p_{\beta}\right) \, ,
\end{eqnarray}
where the pole residues are defined by
\begin{eqnarray}
  \langle 0|J_\mu(0)|Y^-(p)\rangle &=&\lambda_{Y_{-}}\varepsilon_\mu\, , \nonumber\\
  \langle 0|\tilde{J}_{\mu\nu}(0)|Y^-(p)\rangle &=& \frac{\lambda_{Y_{-}}}{M_{Y_{-}}} \, \varepsilon_{\mu\nu\alpha\beta} \, \varepsilon^{\alpha}p^{\beta}\, , \nonumber\\
 \langle 0|\tilde{J}_{\mu\nu}(0)|Z^+(p)\rangle &=&\frac{\lambda_{Z_{+}}}{M_{Z_{+}}} \left(\varepsilon_{\mu}p_{\nu}-\varepsilon_{\nu}p_{\mu} \right)\, , \nonumber\\
  \langle 0|J_{-,\mu\nu}^{AA}(0)|Z^+(p)\rangle &=& \frac{\lambda_{Z_{+}}}{M_{Z_{+}}} \, \varepsilon_{\mu\nu\alpha\beta} \, \varepsilon^{\alpha}p^{\beta}\, , \nonumber\\
 \langle 0|J_{-,\mu\nu}^{AA}(0)|Y^-(p)\rangle &=&\frac{\lambda_{Y_{-}}}{M_{Y_{-}}} \left(\varepsilon_{\mu}p_{\nu}-\varepsilon_{\nu}p_{\mu} \right)\, ,
\end{eqnarray}
the  $\varepsilon_{\mu/\alpha}$ are the polarization vectors, we use the superscripts/subscripts $\pm$  to represent  the positive and negative parity, respectively, and  introduce the notation $Z$ to represent the tetraquark states with the positive parity. We choose the components $\Pi_{-}(p^2)$ and $p^2\widetilde{\Pi}_{-}(p^2)$ to investigate  the  hidden-charm-hidden-strange tetraquark states with the $J^{PC}=1^{--}$ and $1^{-+}$, respectively, so as not to get polluted QCD sum rules.

In fact, there also exist  two-meson scattering state contributions, if they have the same quantum numbers as the  intermediate tetraquark states, because  the quantum field theory does not forbid the current-two-meson couplings. In the QCD sum rules, we take the local currents, the  four  valence  quarks lie almost at the same point, which requires  that the four-quark states, irrespective of the color antitriplet-triplet type or singlet-singlet type,  have the average
spatial sizes  of the same magnitude as the traditional  mesons.  The physical mesons with two valence quarks are spatial extended objects and have average  spatial sizes $\sqrt{\langle r^2\rangle}\sim 0.5\,\rm{fm}$. If the couplings between the local currents and two-meson scattering states are strong enough,  the
average  spatial sizes of the two-meson scattering states are larger or about $1\,\rm{fm}$, and cannot be interpolated by the local currents in the QCD sum rules. In the local limit, the spatial separations are small enough, the charmed  meson pairs  lose themselves and merge into four-quark states, while the net  effects of the current-two-meson couplings can be safely absorbed into the pole residues \cite{WangHuangtao-2014-PRD,WZG-comment,WZG-Z4200}.

At the QCD side of the $\Pi_{\mu\nu}(p)$ and $\Pi_{\mu\nu\alpha\beta}(p)$, there are  two heavy/light quark propagators.  If each heavy quark line emits a gluon and each light quark line contributes  a quark-antiquark  pair, we obtain a quark-gluon  operator $g_sGg_sG\bar{s}s \bar{s}s$   of dimension 10, therefore we should calculate the  condensates $\langle\bar{s}s\rangle$, $\langle\frac{\alpha_{s}GG}{\pi}\rangle$, $\langle\bar{s}g_{s}\sigma Gs\rangle$, $\langle\bar{s}s\rangle^2$, $g_s^2\langle\bar{s}s\rangle^2$,
$\langle\bar{s}s\rangle \langle\frac{\alpha_{s}GG}{\pi}\rangle$,  $\langle\bar{s}s\rangle  \langle\bar{s}g_{s}\sigma Gs\rangle$,
$\langle\bar{s}g_{s}\sigma Gs\rangle^2$ and $\langle\bar{s}s\rangle^2 \langle\frac{\alpha_{s}GG}{\pi}\rangle$, which are vacuum expectations of the quark-gluon operators of the order $\mathcal{O}(\alpha_s^k)$ with $k\leq 1$ \cite{Wang-tetra-formula,WangHuangtao-2014-PRD,WZG-mole-EPJC-1,WZG-mole-EPJC-2}.  We  recalculate the higher  dimensional condensates with the identity $\frac{1}{4}\lambda^a_{ij}\lambda^a_{mn}=-\frac{1}{6}\delta_{ij}\delta_{mn}+\frac{1}{2}\delta_{jm}\delta_{in}$, where the $\lambda^a$ is the Gell-Mann matrix, and reach slightly  different expressions for the spectral densities of the  $J^{PA}_{-,\mu}(x)$, $J^{PA}_{+,\mu}(x)$ and $J^{SV}_{-,\mu}(x)$ \cite{WangY4360Y4660-1803,Wang-tetra-formula}.     The condensate $g_s^2\langle \bar{s}s\rangle^2$ comes from the matrix elements
$\langle \bar{s}\gamma_\mu t^a s g_s D_\eta G^a_{\lambda\tau}\rangle$, $\langle\bar{s}_jD^{\dagger}_{\mu}D^{\dagger}_{\nu}D^{\dagger}_{\alpha}s_i\rangle$  and
$\langle\bar{s}_jD_{\mu}D_{\nu}D_{\alpha}s_i\rangle$, rather than  come from the radiative corrections for the  condensate $\langle \bar{s}s\rangle^2$, where $D_\alpha=\partial_\alpha-ig_sG^n_\alpha \frac{\lambda^n}{2}$,  its contributions are very small and are neglected in most of the QCD sum rules.

We match  the hadron side with the QCD  side of the components $\Pi_{-}(p^2)$ and $p^2\widetilde{\Pi}_{-}(p^2)$ below the continuum thresholds  $s_0$, and perform Borel transform  in regard  to the variable
 $P^2=-p^2$ to acquire the QCD sum rules:
\begin{eqnarray}\label{QCDSR}
\lambda^2_{Y}\, \exp\left(-\frac{M^2_{Y}}{T^2}\right)= \int_{4m_c^2}^{s_0} ds\, \rho(s) \, \exp\left(-\frac{s}{T^2}\right) \, ,
\end{eqnarray}
the lengthy QCD spectral densities $\rho(s)$ are neglected.

We differentiate  Eq.\eqref{QCDSR} in regard  to the variable  $\tau=\frac{1}{T^2}$,  and obtain the QCD sum rules for  the masses of the  hidden-charm-hidden-strange  tetraquark states $Y$,
 \begin{eqnarray}\label{mass-QCDSR}
 M^2_{Y}&=& -\frac{\int_{4m_c^2}^{s_0} ds\frac{d}{d \tau}\rho(s)\exp\left(-\tau s \right)}{\int_{4m_c^2}^{s_0} ds \rho(s)\exp\left(-\tau s\right)}\, .
\end{eqnarray}

\section{Numerical results and discussions}
The quark masses and vacuum condensates depend on the energy scale,
\begin{eqnarray}
\langle\bar{s}s \rangle(\mu)&=&\langle\bar{s}s \rangle({\rm 1GeV})\left[\frac{\alpha_{s}({\rm 1GeV})}{\alpha_{s}(\mu)}\right]^{\frac{12}{33-2n_f}}\, , \nonumber\\
 \langle\bar{s}g_s \sigma Gs \rangle(\mu)&=&\langle\bar{s}g_s \sigma Gs \rangle({\rm 1GeV})\left[\frac{\alpha_{s}({\rm 1GeV})}{\alpha_{s}(\mu)}\right]^{\frac{2}{33-2n_f}}\, , \nonumber\\
 m_c(\mu)&=&m_c(m_c)\left[\frac{\alpha_{s}(\mu)}{\alpha_{s}(m_c)}\right]^{\frac{12}{33-2n_f}} \, ,\nonumber\\
m_s(\mu)&=&m_s({\rm 2GeV} )\left[\frac{\alpha_{s}(\mu)}{\alpha_{s}({\rm 2GeV})}\right]^{\frac{12}{33-2n_f}}\, ,\nonumber\\
\alpha_s(\mu)&=&\frac{1}{b_0t}\left[1-\frac{b_1}{b_0^2}\frac{\log t}{t} +\frac{b_1^2(\log^2{t}-\log{t}-1)+b_0b_2}{b_0^4t^2}\right]\, ,
\end{eqnarray}
 where $t=\log \frac{\mu^2}{\Lambda_{QCD}^2}$, $b_0=\frac{33-2n_f}{12\pi}$, $b_1=\frac{153-19n_f}{24\pi^2}$, $b_2=\frac{2857-\frac{5033}{9}n_f+\frac{325}{27}n_f^2}{128\pi^3}$,  $\Lambda_{QCD}=210\,\rm{MeV}$, $292\,\rm{MeV}$  and  $332\,\rm{MeV}$ for the flavors  $n_f=5$, $4$ and $3$, respectively  \cite{PDG,Narison-mix}. We take the flavor $n_f=4$ as we study the $cs\bar{c}\bar{s}$ states.
 At the initial points, we take  the traditional  values of the condensates $\langle
\bar{q}q \rangle=-(0.24\pm 0.01\, \rm{GeV})^3$,  $\langle\bar{s}s \rangle=(0.8\pm0.1)\langle\bar{q}q \rangle$, $\langle\bar{s}g_s\sigma G s \rangle=m_0^2\langle \bar{s}s \rangle$,
$m_0^2=(0.8 \pm 0.1)\,\rm{GeV}^2$,  $\langle \frac{\alpha_s
GG}{\pi}\rangle=(0.012\pm0.004)\,\rm{GeV}^4 $    at the energy scale  $\mu=1\, \rm{GeV}$
\cite{SVZ79-1,SVZ79-2,Reinders85,Colangelo-Review}, and take the $\overline{MS}$ mass $m_{c}(m_c)=(1.275\pm0.025)\,\rm{GeV}$ and $m_s(\mu=2\,\rm{GeV})=0.095\pm 0.005\,\rm{GeV}$ from The Review of Particle Physics \cite{PDG}.

In this work, we use the energy scale formula $\mu=\sqrt{M^2_{X/Y/Z}-(2{\mathbb{M}}_c)^2}-2{\mathbb{M}}_s$ to determine the best energy scales of the QCD spectral densities \cite{Wang-tetra-formula,WangEPJC-1601}, where the ${\mathbb{M}}_c$ and ${\mathbb{M}}_s$ are the effective $c$ and $s$-quark masses, respectively. We take  the updated value  ${\mathbb{M}}_c=1.82\,\rm{GeV}$ \cite{Wang-tetra-formula,WangEPJC-1601}, and take account of the light-flavor $SU(3)$-breaking effects by introducing an  effective $s$-quark mass ${\mathbb{M}}_s=0.12\,\rm{GeV}$ for the P-wave states, while for the S-wave states, ${\mathbb{M}}_s=0.20\,\rm{GeV}$ \cite{WZG-mole-IJMPA,WZG-XQ-mole-EPJA,WZG-tetra-Zcs-CPC}.
 The energy scale formula  can  enhance  the pole contributions remarkably and improve the convergent behaviors of the operator product expansion remarkably, therefore it is more easier  to acquire  the lowest  tetraquark  (molecule) masses.

In the picture of hidden-charm tetraquark states, we can tentatively assign the $X(3915)$ and $X(4500)$ as the 1S and 2S  states with the $J^{PC}=0^{++}$ \cite{X4140-tetraquark-Lebed,X3915-X4500-EPJC-WZG}, assign
the $Z_c(3900)$ and $Z_c(4430)$   as  the 1S and 2S  states with the $J^{PC}=1^{+-}$, respectively \cite{Maiani-II-type,Nielsen-1401,WangZG-Z4430-CTP},   assign the $Z_c(4020)$ and $Z_c(4600)$ as the 1S and 2S states with the $J^{PC}=1^{+-}$, respectively  \cite{ChenHX-Z4600-A,WangZG-axial-Z4600}, and assign the $X(4140)$ and $X(4685)$ as the 1S and 2S  states with the $J^{PC}=1^{++}$, respectively \cite{WZG-Di-X4140-EPJC,WZG-X4140-X4685}. The energy  gaps between the 1S and 2S states  are about $0.57\sim 0.59 \,\rm{GeV}$. In Ref.\cite{WZG-tetra-vector-NPB}, we  choose the continuum threshold parameters as $\sqrt{s_0}=M_Y+0.4\sim0.6\,\rm{GeV}$ for the hidden-charm tetraquark states with the $J^{PC}=1^{--}$ and
$1^{-+} $. Now we choose  slightly larger parameters $\sqrt{s_0}=M_Y+0.50\sim0.55\pm 0.10\,\rm{GeV}$ so as to include the ground state contributions as large as possible, and perform a consistent and detailed analysis. The relation  $\sqrt{s_0}=M_Y+0.50\sim0.55\pm 0.10\,\rm{GeV}$ serves as a strict constraint.

We search for the best Borel parameters and continuum threshold parameters via trial and error to satisfy
the pole dominance and convergence of the operator product expansion.
The pole contributions (PC)  are defined by
\begin{eqnarray}
{\rm{PC}}&=&\frac{\int_{4m_{c}^{2}}^{s_{0}}ds\rho\left(s\right)\exp\left(-\frac{s}{T^{2}}\right)} {\int_{4m_{c}^{2}}^{\infty}ds\rho\left(s\right)\exp\left(-\frac{s}{T^{2}}\right)}\, ,
\end{eqnarray}
 and the contributions of the  condensates of dimension $n$ are defined by
\begin{eqnarray}
D(n)&=&\frac{\int_{4m_{c}^{2}}^{s_{0}}ds\rho_{n}(s)\exp\left(-\frac{s}{T^{2}}\right)}
{\int_{4m_{c}^{2}}^{s_{0}}ds\rho\left(s\right)\exp\left(-\frac{s}{T^{2}}\right)}\, .
\end{eqnarray}

 Finally,  we obtain the Borel windows, continuum threshold parameters, suitable energy scales of the QCD spectral densities and  pole contributions, which are shown explicitly in Table \ref{BorelP}.
From the Table,  we can see clearly that the pole contributions are about $(40-60)\%$,  the central values are larger than $50\%$, the pole dominance is well satisfied.
In calculations, we observe  that the main contributions come from the perturbative terms, the higher dimensional condensates play a minor important role, just like that in Ref.\cite{WZG-tetra-vector-NPB}, and the contributions $|D(10)|\ll 1\%$, the operator product  expansion converges  very good. It is reliable to extract the masses and pole residues in the Borel windows.

We take  account of all the uncertainties of the relevant  parameters and obtain the masses and pole residues of the hidden-charm-hidden-strange   tetraquark states with the $J^{PC}=1^{--}$ and $1^{-+}$, which are  shown explicitly in Table \ref{mass-residue}. From  Tables \ref{BorelP}--\ref{mass-residue}, we can observe  that the modified energy scale formula $\mu=\sqrt{M^2_{X/Y/Z}-(2{\mathbb{M}}_c)^2}-2{\mathbb{M}}_s$ is  satisfied very good and the relation $\sqrt{s_0}=M_Y+0.50\sim0.55\pm 0.10\,\rm{GeV}$ is also  satisfied very good.

For the conventional charmonium and bottomonium states $\eta_c$, $J/\psi$, $\eta_b$,
$\Upsilon$, $\chi_{b0}$, $\chi_{b1}$, $h_{b}$, $\chi_{b2}$ and $B_c$, the
energy gaps between the ground state and first radial excited state are $654\,\rm{MeV}$,
$589\,\rm{MeV}$, $600\,\rm{MeV}$, $563\,\rm{MeV}$, $373\,\rm{MeV}$, $363\,\rm{MeV}$, $361\,\rm{MeV}$, $356\,\rm{MeV}$ and $597\,\rm{MeV}$, respectively, from The Review of Particle Physics \cite{PDG}.
We can see clearly that the energy gaps are about $0.6\,\rm{GeV}$ for the charmonium states. On the other hand, the energy gap between
the pseudoscalar charm-strange  mesons $D_s$ and $D_{s}(2590)$ is $622\,\rm{MeV}$ from The Review of Particle Physics \cite{PDG}. We estimate that the energy gaps of the ground state and first radial excited state of the $cs\bar{c}\bar{s}$ states are also about $0.6\,\rm{GeV}$, just like the pseudoscalar charm-strange  mesons or charmonium states. If there are some contaminations from the first radial excited states, they are suppressed by the factor
$\exp\left(-\frac{s_0}{T^2} \right)<0.0001\sim0.002$, which is small enough.

\begin{figure}
 \centering
  \includegraphics[totalheight=5cm,width=7cm]{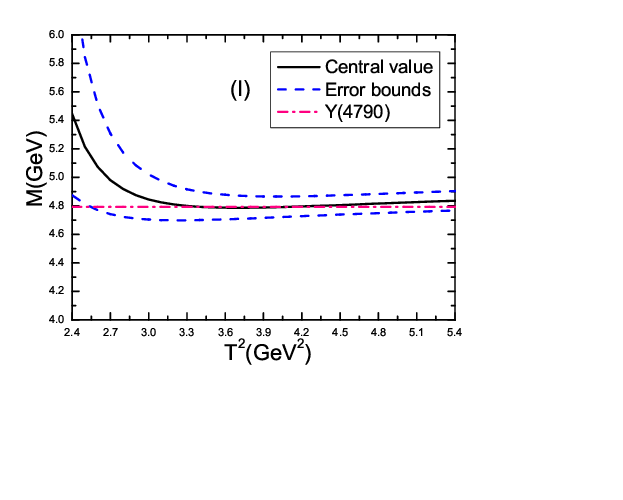}
  \includegraphics[totalheight=5cm,width=7cm]{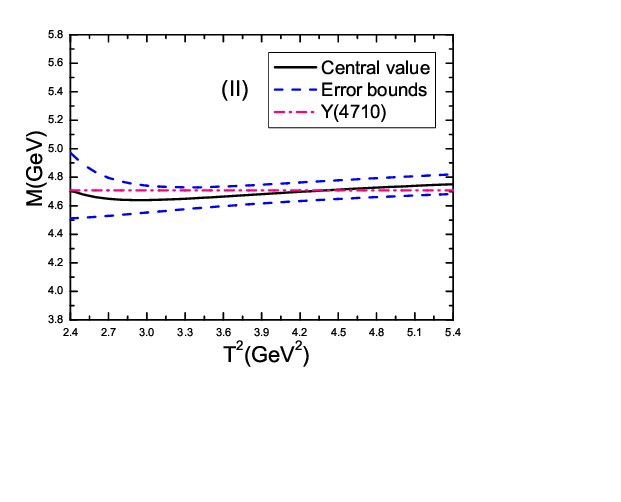}
  \includegraphics[totalheight=5cm,width=7cm]{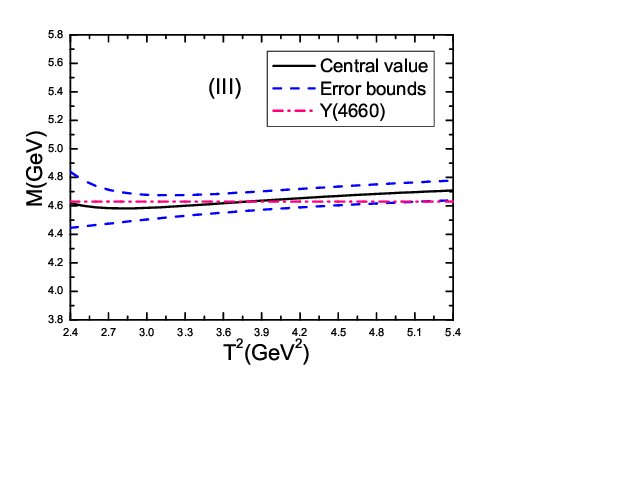}
  \includegraphics[totalheight=5cm,width=7cm]{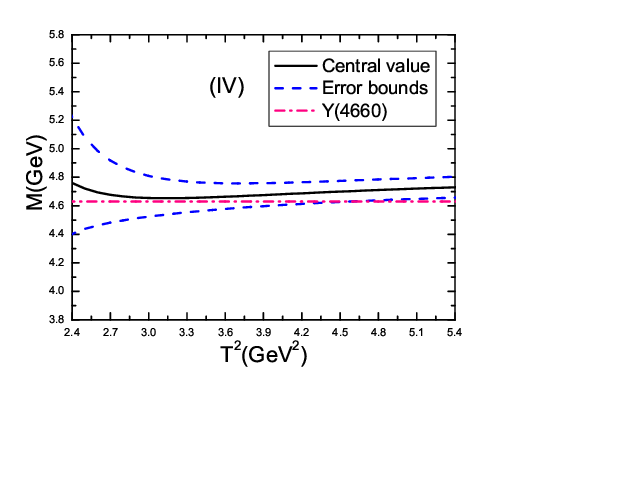}
        \caption{The masses of the hidden-charm-hidden-strange  tetraquark states with the $J^{PC}=1^{--}$   via the  Borel  parameters, where the (I), (II), (III) and (IV)  correspond to   the $[sc]_{P}[\overline{sc}]_{A}-[sc]_{A}[\overline{sc}]_{P}$,
 $[sc]_{\tilde{V}}[\overline{sc}]_{A}-[sc]_{A}[\overline{sc}]_{\tilde{V}}$,
  $[sc]_{\tilde{A}}[\overline{sc}]_{V}+[sc]_{V}[\overline{sc}]_{\tilde{A}}$
    and $[sc]_{S}[\overline{sc}]_{\tilde{V}}-[sc]_{\tilde{V}}[\overline{sc}]_{S}$ tetraquark states,  respectively.}\label{mass-1-fig}
\end{figure}

 In  Fig.\ref{mass-1-fig}, we plot the masses of the  $[sc]_{P}[\overline{sc}]_{A}-[sc]_{A}[\overline{sc}]_{P}$,
 $[sc]_{\tilde{V}}[\overline{sc}]_{A}-[sc]_{A}[\overline{sc}]_{\tilde{V}}$,
  $[sc]_{\tilde{A}}[\overline{sc}]_{V}+[sc]_{V}[\overline{sc}]_{\tilde{A}}$
    and $[sc]_{S}[\overline{sc}]_{\tilde{V}}-[sc]_{\tilde{V}}[\overline{sc}]_{S}$ tetraquark states with the $J^{PC}=1^{--}$  with variations of the Borel parameters for example. From the figure, we observe  that there appear very flat platforms in the Borel windows indeed, the QCD sum rules work well. In Fig.\ref{mass-1-fig}, we also present the experimental  masses of the $Y(4790)$, $Y(4710)$ and $Y(4660)$ \cite{Y4790-BESIII,Y4710-BESIII,PDG}. From the figure, we can see clearly that it is reasonable to assign the $Y(4790)$, $Y(4710)$ and $Y(4660)$ as the
$[sc]_{P}[\overline{sc}]_{A}-[sc]_{A}[\overline{sc}]_{P}$,
$[sc]_{\tilde{V}}[\overline{sc}]_{A}-[sc]_{A}[\overline{sc}]_{\tilde{V}}$ and
$[sc]_{\tilde{A}}[\overline{sc}]_{V}+[sc]_{V}[\overline{sc}]_{\tilde{A}}$ (or $[sc]_{S}[\overline{sc}]_{\tilde{V}}-[sc]_{\tilde{V}}[\overline{sc}]_{S}$) tetraquark states with the $J^{PC}=1^{--}$, respectively, also see Table \ref{Assignments-Table-ss}.

In Table \ref{Assignments-Table}, we present the possible assignments of the  hidden-charm tetraquark states with the $J^{PC}=1^{--}$ and $1^{-+}$ obtained in Ref.\cite{WZG-tetra-vector-NPB}, which  support assigning  the $Y(4360/4390)$ to be the $[uc]_{S}[\overline{uc}]_{V}+[dc]_{S}[\overline{dc}]_{V}+[uc]_{V}[\overline{uc}]_{S}+[dc]_{V}[\overline{dc}]_{S}$ tetraquark state with the  $J^{PC}=1^{--}$ and assigning the
$Y(4660)$ to be the $[uc]_{P}[\overline{uc}]_{A}+[dc]_{P}[\overline{dc}]_{A}-[uc]_{A}[\overline{uc}]_{P}-[dc]_{A}[\overline{dc}]_{P}$  or  $[uc]_{A}[\overline{uc}]_{A}+[dc]_{A}[\overline{dc}]_{A}$  tetraquark state with the  $J^{PC}=1^{--}$.

In 2021, the LHCb collaboration observed the $X(4630)$ in the $J/\psi\phi$ invariant mass spectrum, the measured
Breit-Wigner mass and width are  $4626 \pm 16 {}^{+18}_{-110}\,\rm{MeV}$ and $174 \pm 27 {}^{+134}_{-73}\,\rm{MeV}$, respectively \cite{X4630-LHCb}. Considering the large uncertainties, it is possible to assign the
$X(4630)$ as the $[sc]_{S}[\overline{sc}]_{\tilde{V}}+[sc]_{\tilde{V}}[\overline{sc}]_{S}$ state with the   $J^{PC}=1^{-+}$, which has a mass $4.68\pm0.09\, \rm{GeV}$, see Table \ref{Assignments-Table-ss}.
In Ref.\cite{WZG-Landau}, we prove that it is feasible and reliable to investigate  the multiquark states in the framework of  the QCD sum rules, and  obtain the prediction for  the mass of the $D_s^*\bar{D}_{s1}-D_{s1}\bar{D}_s^*$ molecular state with the exotic quantum numbers  $J^{PC}=1^{-+}$,  $M_{X}=4.67\pm0.08\,\rm{GeV}$, which was obtained before the LHCb data and is compatible with the LHCb data. The $X(4630)$ maybe have two important Fock components.

If only the mass is concerned, the $Y(4660)$ can be assigned as the $[sc]_{\tilde{A}}[\overline{sc}]_{V}+[sc]_{V}[\overline{sc}]_{\tilde{A}}$, $[sc]_{S}[\overline{sc}]_{\tilde{V}}-[sc]_{\tilde{V}}[\overline{sc}]_{S}$, $[uc]_{P}[\overline{uc}]_{A}+[dc]_{P}[\overline{dc}]_{A}-[uc]_{A}[\overline{uc}]_{P}-[dc]_{A}[\overline{dc}]_{P}$  or  $[uc]_{A}[\overline{uc}]_{A}+[dc]_{A}[\overline{dc}]_{A}$ tetraquark state, see Tables \ref{Assignments-Table-ss}-\ref{Assignments-Table}. In other words, the $Y(4660)$ maybe have several important Fock components, we have to study the strong decays in details to diagnose its nature. For example, if we assign $Y(4660)=[sc]_{\tilde{A}}[\overline{sc}]_{V}+[sc]_{V}[\overline{sc}]_{\tilde{A}}$ or $[sc]_{S}[\overline{sc}]_{\tilde{V}}-[sc]_{\tilde{V}}[\overline{sc}]_{S}$, then the
strong decays $Y(4660)\to J/\psi f_0(980)$, $ \eta_c \phi$,    $ \chi_{c0}\phi$, $ D_s \bar{D}_s$, $ D_s^* \bar{D}^*_s$, $ D_s \bar{D}^*_s$,  $ D_s^* \bar{D}_s$, $  J/\psi \pi^+\pi^-$ and $  \psi^\prime \pi^+\pi^-$ are Okubo-Zweig-Iizuka super-allowed, considering the intermediate process  $f_0(980)\to \pi^+\pi^-$. Up to now, only the decays $Y(4660)\to J/\psi \pi^+\pi^-$, $\psi_2(3823)\pi^+\pi^-$, $\Lambda_c^+\Lambda_c^-$ and $D^+_sD^{-}_{s1}$ have been observed \cite{PDG}, which cannot exclude the assignments $Y(4660)=[uc]_{P}[\overline{uc}]_{A}+[dc]_{P}[\overline{dc}]_{A}-[uc]_{A}[\overline{uc}]_{P}-[dc]_{A}[\overline{dc}]_{P}$  or  $[uc]_{A}[\overline{uc}]_{A}+[dc]_{A}[\overline{dc}]_{A}$, as the decay $Y(4660)\to D^+_sD^{-}_{s1}$ can take place through the re-scattering mechanism.  More theoretical works and more experimental data are still needed.

After Ref.\cite{WZG-tetra-vector-NPB} has been published, the $Y(4500)$ was observed by the BESIII collaboration \cite{BESIII-KK-4500,X4500-BESIII}.  At the energy about $4.5\,\rm{GeV}$, we obtain three hidden-charm tetraquark states with the $J^{PC}=1^{--}$, the $[uc]_{\tilde{V}}[\overline{uc}]_{A}+[dc]_{\tilde{V}}[\overline{dc}]_{A}
-[uc]_{A}[\overline{uc}]_{\tilde{V}}-[dc]_{A}[\overline{dc}]_{\tilde{V}}$,
 $[uc]_{\tilde{A}}[\overline{uc}]_{V}+[dc]_{\tilde{A}}[\overline{dc}]_{V}
 +[uc]_{V}[\overline{uc}]_{\tilde{A}}+[dc]_{V}[\overline{dc}]_{\tilde{A}}$ and
 $[uc]_{S}[\overline{uc}]_{\tilde{V}}+[dc]_{S}[\overline{dc}]_{\tilde{V}}
 -[uc]_{\tilde{V}}[\overline{uc}]_{S}-[dc]_{\tilde{V}}[\overline{dc}]_{S}$ tetraquark states have
 the masses $4.53\pm0.07\, \rm{GeV}$, $4.48\pm0.08\,\rm{GeV}$ and $4.50\pm0.09\,\rm{GeV}$, respectively \cite{WZG-tetra-vector-NPB}.
  In Ref.\cite{Y4500-decay-NPB}, we take the $Y(4500)$ as the  $[uc]_{\tilde{A}}[\overline{uc}]_{V}+[dc]_{\tilde{A}}[\overline{dc}]_{V}
  +[uc]_{V}[\overline{uc}]_{\tilde{A}}+[dc]_{V}[\overline{dc}]_{\tilde{A}}$ tetraquark state, and  study the three-body strong decay $Y(4500)\to D^{*-}D^{*0}\pi^+$ with the light-cone QCD sum rules.

In Ref.\cite{Vector-Tetra-WZG-P-wave}, we introduce an explicit P-wave to construct the four-quark currents to study the  hidden-charm tetraquark states with the $J^{PC}=1^{--}$,  and the calculations support   assigning the $Y(4260/4220)$  as the  $C\gamma_5\otimes\stackrel{\leftrightarrow}{\partial}_\mu\otimes \gamma_5C$-type tetraquark state;
 assigning the $Y(4260/4220)$ or $Y(4360/4320)$ as the  $C\gamma_\alpha\otimes\stackrel{\leftrightarrow}{\partial}_\mu\otimes \gamma^{\alpha}C$   type  tetraquark state;  assigning the $Y(4360/4320)$ or $Y(4390)$ as the
 $C\gamma_\mu \otimes\stackrel{\leftrightarrow}{\partial}_\alpha \otimes\gamma^\alpha C +C\gamma^\alpha \otimes\stackrel{\leftrightarrow}{\partial}_\alpha \otimes\gamma_\mu C$ type or $C\gamma_5 \otimes\stackrel{\leftrightarrow}{\partial}_\mu \otimes\gamma_\nu C
+C\gamma_\nu \otimes\stackrel{\leftrightarrow}{\partial}_\mu \otimes\gamma_5 C
-C\gamma_5 \otimes\stackrel{\leftrightarrow}{\partial}_\nu \otimes\gamma_\mu C
-C\gamma_\mu \otimes\stackrel{\leftrightarrow}{\partial}_\nu \otimes\gamma_5 C $ type  tetraquark state.

All in all, there are enough rooms to accommodate the existing $Y$ states above $4.2\,\rm{GeV}$. However, due to uncertainties of the predictions, we cannot distinguish the quark structures based on the masses alone in the current accuracy, comprehensive and detailed studies on the strong decays are still needed. Again, for the  example, at the energy about $4.5\,\rm{GeV}$, there
exist three hidden-charm tetraquark states with the $J^{PC}=1^{--}$ and Isospin $I=0$, i.e. $Y=$ $[uc]_{\tilde{V}}[\overline{uc}]_{A}+[dc]_{\tilde{V}}[\overline{dc}]_{A}
-[uc]_{A}[\overline{uc}]_{\tilde{V}}-[dc]_{A}[\overline{dc}]_{\tilde{V}}$,
 $[uc]_{\tilde{A}}[\overline{uc}]_{V}+[dc]_{\tilde{A}}[\overline{dc}]_{V}
 +[uc]_{V}[\overline{uc}]_{\tilde{A}}+[dc]_{V}[\overline{dc}]_{\tilde{A}}$ or
 $[uc]_{S}[\overline{uc}]_{\tilde{V}}+[dc]_{S}[\overline{dc}]_{\tilde{V}}
 -[uc]_{\tilde{V}}[\overline{uc}]_{S}-[dc]_{\tilde{V}}[\overline{dc}]_{S}$ \cite{WZG-tetra-vector-NPB}. We have  studied the strong decays
 $Y \to D\bar{D}$, $D^*\bar{D}^*$, $D\bar{D}^*$, $D^*\bar{D}$, $\omega J/\psi $, $\omega \eta_c $, $\omega\chi_{c0}$, $\omega\chi_{c1}$, $J/\psi f_0(500)$ with the three-point QCD sum rules in a systematic way, and obtained  very interesting results, which would be presented in another work after carefully checking. By precisely measuring the ratios among the
 partial decay widths, we can distinguish the quark structures. Analogously, we can study the strong decays
 $Y \to D_s\bar{D}_s$, $D_s^*\bar{D}_s^*$, $D_s\bar{D}_s^*$, $D_s^*\bar{D}_s$, $\phi J/\psi $, $\phi \eta_c $, $\phi\chi_{c0}$, $\phi\chi_{c1}$, $J/\psi f_0(980)$ in the present case.

We can confront those hidden-charm-hidden-strange  tetraquark states  to the experimental data  at the BESIII, LHCb, Belle II,  CEPC, FCC, ILC
 in the future, which maybe play an important role in interpreting  the exotic $X$, $Y$, $Z$ states.
 For example, we can search for the  $Y$  states with the $J^{PC}=1^{--}$ and $1^{-+}$ in the two-body or three-body strong decays,
\begin{eqnarray}
  Y(1^{--}) &\to&  \chi_{c0}\rho/\omega/\phi \, ,\, J/\psi \pi^+\pi^- \, ,\,   J/\psi K\bar{K}\, ,\,  \eta_c\rho/\omega/\phi\, ,\, \chi_{c1}\rho/\omega/\phi\, , \nonumber\\
  Y(1^{-+}) &\to& J/\psi\rho/\omega/\phi \, ,\, h_c\rho/\omega/\phi \, ,
\end{eqnarray}
just as what have been done previously.

\begin{table}
\begin{center}
\begin{tabular}{|c|c|c|c|c|c|c|c|c|}\hline\hline
 $Y_c$                                                                     &$J^{PC}$  &$T^2(\rm{GeV}^2)$ &$\sqrt{s_0}(\rm GeV) $ &$\mu(\rm{GeV})$  &pole           \\ \hline

$[sc]_{P}[\overline{sc}]_{A}-[sc]_{A}[\overline{sc}]_{P}$                  &$1^{--}$  &$4.1-4.7$          &$5.35\pm0.10$          &$2.9$            &$(40-61)\%$  \\

$[sc]_{P}[\overline{sc}]_{A}+[sc]_{A}[\overline{sc}]_{P}$                  &$1^{-+}$  &$4.0-4.6$          &$5.30\pm0.10$          &$2.8$            &$(41-61)\%$ \\

$[sc]_{S}[\overline{sc}]_{V}+[sc]_{V}[\overline{sc}]_{S}$                  &$1^{--}$  &$3.5-4.0$          &$5.05\pm0.10$          &$2.5$            &$(41-62)\%$  \\

$[sc]_{S}[\overline{sc}]_{V}-[sc]_{V}[\overline{sc}]_{S}$                  &$1^{-+}$  &$4.0-4.6$          &$5.35\pm0.10$          &$2.9$            &$(40-60)\%$ \\

$[sc]_{\tilde{V}}[\overline{sc}]_{A}-[sc]_{A}[\overline{sc}]_{\tilde{V}}$  &$1^{--}$  &$3.9-4.5$          &$5.25\pm0.10$          &$2.7$            &$(40-61)\%$  \\

$[sc]_{\tilde{V}}[\overline{sc}]_{A}+[sc]_{A}[\overline{sc}]_{\tilde{V}}$  &$1^{-+}$  &$4.0-4.6$          &$5.35\pm0.10$          &$2.9$            &$(40-61)\%$ \\

$[sc]_{\tilde{A}}[\overline{sc}]_{V}+[sc]_{V}[\overline{sc}]_{\tilde{A}}$  &$1^{--}$  &$3.8-4.4$          &$5.20\pm0.10$          &$2.7$            &$(40-61)\%$    \\

$[sc]_{\tilde{A}}[\overline{sc}]_{V}-[sc]_{V}[\overline{sc}]_{\tilde{A}}$  &$1^{-+}$  &$3.9-4.5$          &$5.25\pm0.10$          &$2.7$            &$(40-61)\%$ \\

$[sc]_{S}[\overline{sc}]_{\tilde{V}}-[sc]_{\tilde{V}}[\overline{sc}]_{S}$  &$1^{--}$  &$3.7-4.2$          &$5.20\pm0.10$          &$2.7$            &$(41-62)\%$    \\

$[sc]_{S}[\overline{sc}]_{\tilde{V}}+[sc]_{\tilde{V}}[\overline{sc}]_{S}$  &$1^{-+}$  &$3.7-4.3$          &$5.20\pm0.10$          &$2.7$            &$(40-62)\%$ \\

$[sc]_{P}[\overline{sc}]_{\tilde{A}}-[sc]_{\tilde{A}}[\overline{sc}]_{P}$  &$1^{--}$  &$4.1-4.7$          &$5.30\pm0.10$          &$2.8$            &$(40-60)\%$    \\

$[sc]_{P}[\overline{sc}]_{\tilde{A}}+[sc]_{\tilde{A}}[\overline{sc}]_{P}$  &$1^{-+}$  &$4.1-4.7$          &$5.30\pm0.10$          &$2.8$            &$(40-60)\%$    \\

$[sc]_{A}[\overline{sc}]_{A}$                                              &$1^{--}$  &$4.2-4.9$          &$5.40\pm0.10$          &$3.0$            &$(40-60)\%$   \\

\hline\hline
\end{tabular}
\end{center}
\caption{ The Borel parameters, continuum threshold parameters, energy scales of the QCD spectral densities and  pole contributions  for the ground state hidden-charm-hidden-strange tetraquark states. }\label{BorelP}
\end{table}

\begin{table}
\begin{center}
\begin{tabular}{|c|c|c|c|c|c|c|c|c|}\hline\hline
 $Y_c$                                                                     &$J^{PC}$   &$M_Y (\rm{GeV})$   &$\lambda_Y (\rm{GeV}^5) $            \\ \hline

$[sc]_{P}[\overline{sc}]_{A}-[sc]_{A}[\overline{sc}]_{P}$                  &$1^{--}$   &$4.80\pm0.08$      &$(8.97\pm1.09)\times 10^{-2}$   \\

$[sc]_{P}[\overline{sc}]_{A}+[sc]_{A}[\overline{sc}]_{P}$                  &$1^{-+}$   &$4.75\pm0.08$      &$(8.34\pm1.04)\times 10^{-2}$   \\

$[sc]_{S}[\overline{sc}]_{V}+[sc]_{V}[\overline{sc}]_{S}$                  &$1^{--}$   &$4.53\pm0.08$      &$(5.70\pm0.79)\times 10^{-2}$           \\

$[sc]_{S}[\overline{sc}]_{V}-[sc]_{V}[\overline{sc}]_{S}$                  &$1^{-+}$   &$4.83\pm0.09$      &$(8.39\pm1.05)\times 10^{-2}$           \\

$[sc]_{\tilde{V}}[\overline{sc}]_{A}-[sc]_{A}[\overline{sc}]_{\tilde{V}}$  &$1^{--}$   &$4.70\pm0.08$      &$(1.32\pm0.18)\times 10^{-1}$         \\

$[sc]_{\tilde{V}}[\overline{sc}]_{A}+[sc]_{A}[\overline{sc}]_{\tilde{V}}$  &$1^{-+}$   &$4.81\pm0.09$      &$(1.43\pm0.19)\times 10^{-1}$         \\

$[sc]_{\tilde{A}}[\overline{sc}]_{V}+[sc]_{V}[\overline{sc}]_{\tilde{A}}$  &$1^{--}$   &$4.65\pm0.08$      &$(1.23\pm0.17)\times 10^{-1}$     \\

$[sc]_{\tilde{A}}[\overline{sc}]_{V}-[sc]_{V}[\overline{sc}]_{\tilde{A}}$  &$1^{-+}$   &$4.71\pm0.08$      &$(1.34\pm0.17)\times 10^{-1}$     \\

$[sc]_{S}[\overline{sc}]_{\tilde{V}}-[sc]_{\tilde{V}}[\overline{sc}]_{S}$  &$1^{--}$   &$4.68\pm0.09$      &$(6.22\pm0.82)\times 10^{-2}$     \\

$[sc]_{S}[\overline{sc}]_{\tilde{V}}+[sc]_{\tilde{V}}[\overline{sc}]_{S}$  &$1^{-+}$   &$4.68\pm0.09$      &$(6.23\pm0.84)\times 10^{-2}$     \\

$[sc]_{P}[\overline{sc}]_{\tilde{A}}-[sc]_{\tilde{A}}[\overline{sc}]_{P}$  &$1^{--}$   &$4.75\pm0.08$      &$(7.93\pm0.98)\times 10^{-2}$     \\

$[sc]_{P}[\overline{sc}]_{\tilde{A}}+[sc]_{\tilde{A}}[\overline{sc}]_{P}$  &$1^{-+}$   &$4.75\pm0.08$      &$(7.94\pm0.97)\times 10^{-2}$     \\

$[sc]_{A}[\overline{sc}]_{A}$                                              &$1^{--}$   &$4.85\pm0.09$      &$(8.38\pm1.05)\times 10^{-2}$     \\

\hline\hline
\end{tabular}
\end{center}
\caption{ The masses and pole residues of the ground state  hidden-charm-hidden-strange tetraquark states. }\label{mass-residue}
\end{table}

\begin{table}
\begin{center}
\begin{tabular}{|c|c|c|c|c|c|c|c|c|}\hline\hline
 $Y_c$                                                                     &$J^{PC}$   &$M_Y (\rm{GeV})$   &Assignments            \\ \hline

$[sc]_{P}[\overline{sc}]_{A}-[sc]_{A}[\overline{sc}]_{P}$                  &$1^{--}$   &$4.80\pm0.08$    & ? $Y(4790)$ \\

$[sc]_{P}[\overline{sc}]_{A}+[sc]_{A}[\overline{sc}]_{P}$                  &$1^{-+}$   &$4.75\pm0.08$    &   \\

$[sc]_{S}[\overline{sc}]_{V}+[sc]_{V}[\overline{sc}]_{S}$                  &$1^{--}$   &$4.53\pm0.08$    & \\

$[sc]_{S}[\overline{sc}]_{V}-[sc]_{V}[\overline{sc}]_{S}$                  &$1^{-+}$   &$4.83\pm0.09$    &  \\

$[sc]_{\tilde{V}}[\overline{sc}]_{A}-[sc]_{A}[\overline{sc}]_{\tilde{V}}$  &$1^{--}$   &$4.70\pm0.08$    & ? $Y(4710)$ \\

$[sc]_{\tilde{V}}[\overline{sc}]_{A}+[sc]_{A}[\overline{sc}]_{\tilde{V}}$  &$1^{-+}$   &$4.81\pm0.09$    & \\

$[sc]_{\tilde{A}}[\overline{sc}]_{V}+[sc]_{V}[\overline{sc}]_{\tilde{A}}$  &$1^{--}$   &$4.65\pm0.08$    & ? $Y(4660)$     \\

$[sc]_{\tilde{A}}[\overline{sc}]_{V}-[sc]_{V}[\overline{sc}]_{\tilde{A}}$  &$1^{-+}$   &$4.71\pm0.08$    &   \\

$[sc]_{S}[\overline{sc}]_{\tilde{V}}-[sc]_{\tilde{V}}[\overline{sc}]_{S}$  &$1^{--}$   &$4.68\pm0.09$    & ? $Y(4660)$     \\

$[sc]_{S}[\overline{sc}]_{\tilde{V}}+[sc]_{\tilde{V}}[\overline{sc}]_{S}$  &$1^{-+}$   &$4.68\pm0.09$    &  ?? $X(4630)$ \\

$[sc]_{P}[\overline{sc}]_{\tilde{A}}-[sc]_{\tilde{A}}[\overline{sc}]_{P}$  &$1^{--}$   &$4.75\pm0.08$    &    \\

$[sc]_{P}[\overline{sc}]_{\tilde{A}}+[sc]_{\tilde{A}}[\overline{sc}]_{P}$  &$1^{-+}$   &$4.75\pm0.08$    &    \\

$[sc]_{A}[\overline{sc}]_{A}$                                              &$1^{--}$   &$4.85\pm0.09$    &     \\

\hline\hline
\end{tabular}
\end{center}
\caption{ The possible assignments of the hidden-charm-hidden-strange tetraquark states. }\label{Assignments-Table-ss}
\end{table}

\begin{table}
\begin{center}
\begin{tabular}{|c|c|c|c|c|c|c|c|c|}\hline\hline
  $Y_c$                                                                    & $J^{PC}$  & $M_Y (\rm{GeV})$  & Assignments          \\ \hline

$[uc]_{P}[\overline{dc}]_{A}-[uc]_{A}[\overline{dc}]_{P}$                  &$1^{--}$   &$4.66\pm0.07$      & ?\,\,$Y(4660)$          \\

$[uc]_{P}[\overline{dc}]_{A}+[uc]_{A}[\overline{dc}]_{P}$                  &$1^{-+}$   &$4.61\pm0.07$      &                     \\

$[uc]_{S}[\overline{dc}]_{V}+[uc]_{V}[\overline{dc}]_{S}$                  &$1^{--}$   &$4.35\pm0.08$      & ?\,\,$Y(4360/4390)$     \\

$[uc]_{S}[\overline{dc}]_{V}-[uc]_{V}[\overline{dc}]_{S}$                  &$1^{-+}$   &$4.66\pm0.09$      &                     \\

$[uc]_{\tilde{V}}[\overline{dc}]_{A}-[uc]_{A}[\overline{dc}]_{\tilde{V}}$  &$1^{--}$   &$4.53\pm0.07$      &                     ? $Y(4500)$\\

$[uc]_{\tilde{V}}[\overline{dc}]_{A}+[uc]_{A}[\overline{dc}]_{\tilde{V}}$  &$1^{-+}$   &$4.65\pm0.08$      &                     \\

$[uc]_{\tilde{A}}[\overline{dc}]_{V}+[uc]_{V}[\overline{dc}]_{\tilde{A}}$  &$1^{--}$   &$4.48\pm0.08$      &                     ? $Y(4500)$\\

$[uc]_{\tilde{A}}[\overline{dc}]_{V}-[uc]_{V}[\overline{dc}]_{\tilde{A}}$  &$1^{-+}$   &$4.55\pm0.07$      &                     \\

$[uc]_{S}[\overline{dc}]_{\tilde{V}}-[uc]_{\tilde{V}}[\overline{dc}]_{S}$  &$1^{--}$   &$4.50\pm0.09$      &                     ? $Y(4500)$\\

$[uc]_{S}[\overline{dc}]_{\tilde{V}}+[uc]_{\tilde{V}}[\overline{dc}]_{S}$  &$1^{-+}$   &$4.50\pm0.09$      &                     \\

$[uc]_{P}[\overline{dc}]_{\tilde{A}}-[uc]_{\tilde{A}}[\overline{dc}]_{P}$  &$1^{--}$   &$4.60\pm0.07$      &                     \\

$[uc]_{P}[\overline{dc}]_{\tilde{A}}+[uc]_{\tilde{A}}[\overline{dc}]_{P}$  &$1^{-+}$   &$4.61\pm0.08$      &                     \\

$[uc]_{A}[\overline{dc}]_{A}$                                              &$1^{--}$   &$4.69\pm0.08$      & ?\,\,$Y(4660)$      \\
\hline\hline
\end{tabular}
\end{center}
\caption{ The possible assignments of the  hidden-charm tetraquark states, the isospin limit is implied. }\label{Assignments-Table}
\end{table}

\section{Conclusion}
In this work,  we take the scalar, pseudoscalar, axialvector, vector and tensor  diquarks  as the elementary  building blocks  to construct local  four-quark currents  with an implicit P-wave,  and study the mass spectrum of the hidden-charm-hidden-strange tetraquark states with the $J^{PC}=1^{--}$ and $1^{-+}$ in the framework of  the QCD sum rules consistently and comprehensively,  and revisit the assignments of the  $X/Y$ states in the  picture of tetraquark  states. The predictions favor
assigning the $Y(4790)$, $Y(4710)$ and $Y(4660)$ as the
$[sc]_{P}[\overline{sc}]_{A}-[sc]_{A}[\overline{sc}]_{P}$,
$[sc]_{\tilde{V}}[\overline{sc}]_{A}-[sc]_{A}[\overline{sc}]_{\tilde{V}}$ and
$[sc]_{\tilde{A}}[\overline{sc}]_{V}+[sc]_{V}[\overline{sc}]_{\tilde{A}}$ (or $[sc]_{S}[\overline{sc}]_{\tilde{V}}-[sc]_{\tilde{V}}[\overline{sc}]_{S}$) tetraquark states with the $J^{PC}=1^{--}$, respectively. Considering our previous works, we can draw the conclusion tentatively that the $Y(4660)$ can be assigned as the $[sc]_{\tilde{A}}[\overline{sc}]_{V}+[sc]_{V}[\overline{sc}]_{\tilde{A}}$, $[sc]_{S}[\overline{sc}]_{\tilde{V}}-[sc]_{\tilde{V}}[\overline{sc}]_{S}$, $[uc]_{P}[\overline{uc}]_{A}+[dc]_{P}[\overline{dc}]_{A}-[uc]_{A}[\overline{uc}]_{P}-[dc]_{A}[\overline{dc}]_{P}$  or  $[uc]_{A}[\overline{uc}]_{A}+[dc]_{A}[\overline{dc}]_{A}$ tetraquark state, in other words,  the $Y(4660)$ maybe have several important Fock components, we have to study the strong decays  exclusively in details to diagnose its nature.
In summary, we can  assign  all the exotic $Y$ states above $4.2\,\rm{GeV}$ in a consistent way, although the
assignments are not definite in the current accuracy.
Furthermore, the predictions favor assigning the
$X(4630)$ as the $[sc]_{S}[\overline{sc}]_{\tilde{V}}+[sc]_{\tilde{V}}[\overline{sc}]_{S}$ state with the   $J^{PC}=1^{-+}$.
We can confront all the  hidden-charm/hidden-charm-hidden-strange  tetraquark states with the $J^{PC}=1^{--}$ and $1^{-+}$  to the experimental data   in the future.

\section*{Acknowledgements}
This research was supported by the National Natural Science Foundation of
China through Grant No.12175068.

\end{document}